\documentclass[pra,twocolumn,superscriptaddress,nobibnotes,longbibliography,notitlepage,nofootinbib,floatfix]{revtex4-2}

\usepackage{soul}
\usepackage{graphicx}
\usepackage{amssymb,amsmath,color,amsfonts, amsthm}
\usepackage{hyperref}
\hypersetup{
	colorlinks,
	linkcolor={blue},
	citecolor={blue},
	urlcolor={blue}
}
\usepackage{ragged2e}
\usepackage[capitalise]{cleveref}
\usepackage[normalem]{ulem}
\usepackage[utf8]{inputenc}
\usepackage[english]{babel}
\usepackage{units}
\usepackage{braket}
\usepackage[miktex]{gnuplottex}
\usepackage{float}
\usepackage{epstopdf}

\begin{document}

\title{Non-inertial motion dependent entangled Bell-state}

\author{Julius Arthur Bittermann}
\email{Julius.Bittermann@oeaw.ac.at}
\affiliation{Institute for Quantum Optics and Quantum Information -- IQOQI Vienna, Austrian Academy of Sciences, Boltzmanngasse 3, 1090 Vienna, Austria}
\affiliation{Atominstitut,  Technische  Universit{\"a}t  Wien,  Stadionallee 2, 1020  Vienna,  Austria}
\author{Matthias Fink}
\affiliation{Institute for Quantum Optics and Quantum Information -- IQOQI Vienna, Austrian Academy of Sciences, Boltzmanngasse 3, 1090 Vienna, Austria}
\affiliation{present address: Quantum Technology Laboratories GmbH, Clemens-Holzmeister-Stra{\ss}e 6/6, 1100 Vienna, Austria}
\author{Marcus Huber}
\email{marcus.huber@tuwien.ac.at}
\affiliation{Atominstitut,  Technische  Universit{\"a}t  Wien,  Stadionallee 2, 1020  Vienna,  Austria}
\affiliation{Institute for Quantum Optics and Quantum Information -- IQOQI Vienna, Austrian Academy of Sciences, Boltzmanngasse 3, 1090 Vienna, Austria}
\author{Rupert Ursin}
\email{rupert.ursin@qtlabs.at}
\affiliation{Institute for Quantum Optics and Quantum Information -- IQOQI Vienna, Austrian Academy of Sciences, Boltzmanngasse 3, 1090 Vienna, Austria}
\affiliation{present address: Quantum Technology Laboratories GmbH, Clemens-Holzmeister-Stra{\ss}e 6/6, 1100 Vienna, Austria}

\begin{abstract}
We show the targeted phase-manipulation of an entangled photonic Bell state via non-inertial motion. To this end, we place a very compact laboratory, consisting of a SPDC source and a Sagnac interferometer, on a rotating platform (non-inertial reference frame). The photon pairs of a $\ket{\phi}$-state are in a superposition of co- and counter-rotation. The phase of the $\ket{\phi}$-state is linearly dependent on the angular velocity of the rotating platform due to the Sagnac effect. We measure the visibility and certify entanglement with the Bell-CHSH parameter $S$. Additionally, we conduct a partial quantum state tomography on the Bell states in a non-inertial environment. Our experiment showcases the unitary transformation of an entangled state via non-inertial motion and constitutes not only a switch between a $\ket{\phi^{-}}$-state and a $\ket{\phi^{+}}$-state but also a further experiment at the interplay of non-inertial motion and quantum physics.
\end{abstract}

\date{\today}

\maketitle

\noindent\textbf{Introduction}.\ 
Quantum entanglement is one of the most intriguing phenomena of quantum mechanics, describing extremely strong correlations between several particles that have no analogue in classical physics. The creation and manipulation of entanglement is currently undergoing a significant leap from  laboratories to applications, and thus constitutes a key enabling quantum technology. Not only a resource for quantum key distribution \cite{MiciusI,MiciusII} and quantum computing, it also enables new pathways for metrology and allows probing the interplay between non-inertial motion and quantum physics \cite{Fink,Bittermann_2023}. 
In this regard, Mach-Zehnder interferometers in gravitational fields and Sagnac interferometers, are a versatile tool for gravity-related quantum experiments. 
Pioneering examples include neutrons in a Mach-Zehnder interferometer, which show a phase change dependent on the height in a gravitational field \cite{COW1,COW2}. Dependent on the alignment on the surface of the earth these kind of interferometers show also a phase shift due to the Sagnac effect \cite{Werner_1979}. Using modern quantum technologies, the Sagnac effect has now been demonstrated also with single photons \cite{Bertocchi_2006}. 
In a further experiment towards metrological advantages, by using N00N-states with $N=2$ in a Sagnac interferometer, it is shown that the shot noise can be transcended \cite{Fink_2019}. 
These results were surpassed recently, when earth rotation was measured with a Sagnac interferometer with a sensitivity of 5 $\mu$rad/s by using path-entangled states \cite{Silvestri_2023}, almost hundred years after Michelson and Gale measured earth rotation with a Sagnac interferometer for the first time \cite{Michelson_Gale_1925}. 
Moreover, this kind of interferometer enables the targeted manipulation of the distinguishability in time of two photons. As a consequence, the dip of the Hong-Ou-Mandel interference shifts in a rotating and thus non-inertial system, dependent on the angular velocity \cite{HOM_Rot}.
Rotating fiber systems as in \cite{Conc_Rev_Etglm} also enable revealing and concealing of entanglement. Another test with two nested Sagnac interferometers demonstrates how non-inertial motion can be used to switch between a bosonic and a fermionic state \cite{Cromb_2022}.
Originally conceived by George Sagnac in the year 1913 \cite{sagnac1913luminiferous,sagnac1913regarding}, gyroscopes based on the Sagnac-effect are nowadays indispensable for rotation sensing. \\
In this paper, we present a Sagnac-interferometer customized for polarization entangled photon pairs, with which we show how non-inertial motion can affect the phase of a Bell state.  
The easiest way to change a polarized qubit, and thus its position on the Bloch sphere, is the usage of waveplates. Local changes are sufficient to readily switch between four Bell-states, which for example enables a new way of encoding messages, the so-called Quantum Dense Coding \cite{PhysRevLett.76.4656}. Here, the quantum advantage is that two bits of information can be encoded by just applying a single unitary transformation on only one particle. 
In this paper, we use rotational motion in order to change the phase of a Bell state. 

\begin{figure*}[t] 
\centering
\includegraphics[width=1.00\textwidth]{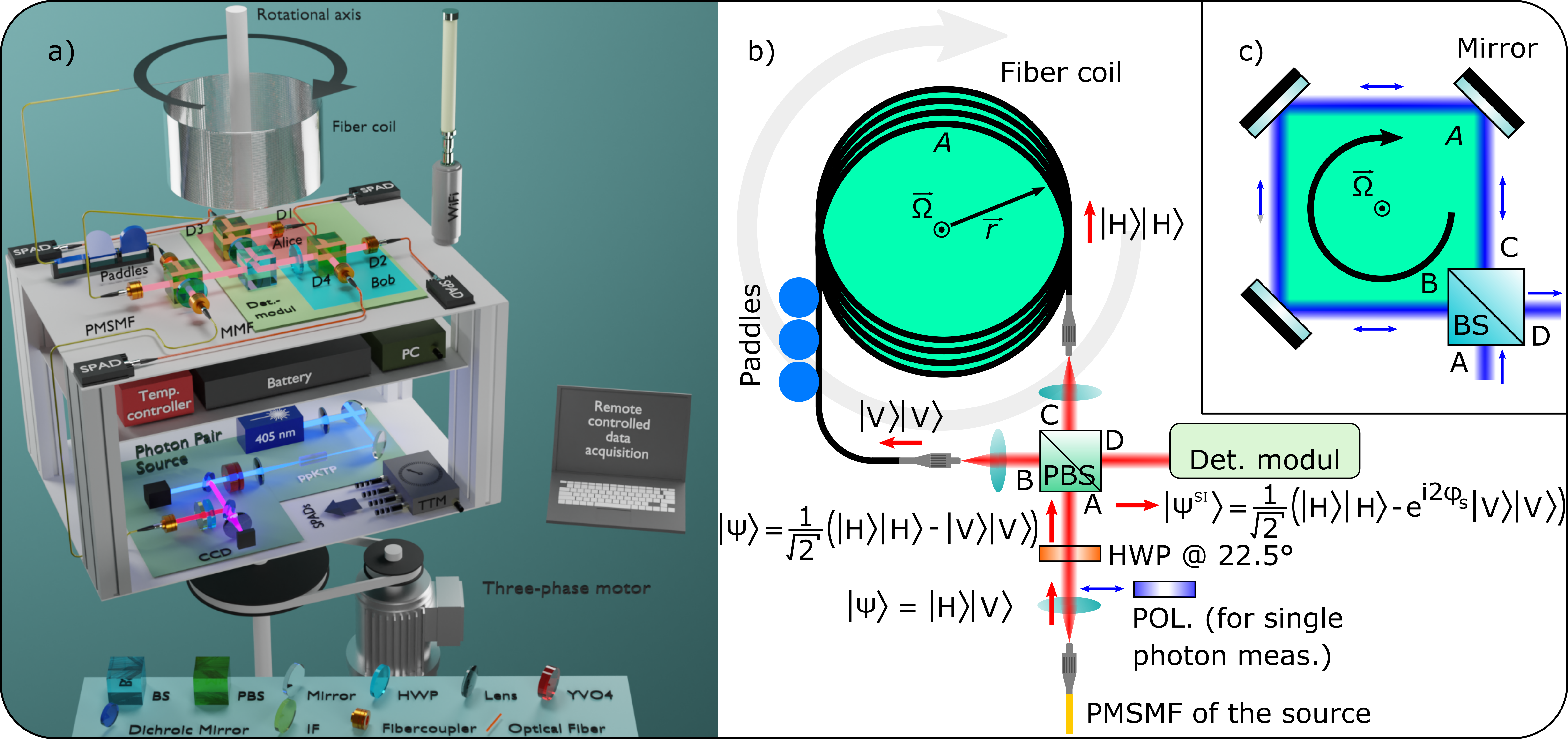}
\caption{\textbf{Illustration of the setup of our Sagnac-interferometer for Bell-states} a) Rotating laboratory consisting of a 4-level crate system. The first level in the bottom accommodates the SPDC source with a type-II ppKTP crystal and the time tagging module (TTM). For the compensation of the temporal walk-off of the $\ket{\mathrm{H}}$- and $\ket{\mathrm{V}}$-photons, we use two neodymium-doped yttrium orthovanadate (YVO4) crystals  and a polarization maintaining single mode fiber (PMSMF) which guides the $\ket{\mathrm{H},\mathrm{V}}$ photon pairs to level 3. Two dichroic mirrors and an interference filter (IF) separate the pump photons from the SPDC photons. A charged coupled device (CCD) serves for the optical alignment of the laser focus inside the crystal.  Dotted lines going out of the TTM indicate the cables which are connected with the single photon avalanche diodes (SPAD). In the second level there are the battery, providing electrical power for 40-45~min, the temperature controller for the crystal and a PC for the data acquisition. The third level contains the detection module (on the green plate) with polarization analyzers of Alice and Bob and the Sagnac interferometer consisting of a polarizing beamsplitter (PBS) and a fiber coil. A beamsplitter (BS) separates 50 \% of the photon pairs. Half-wave plates (HWP) at an angle of $22.5^\circ$ in front of the polarization analyzers of Alice (on the red plate) and Bob (on the blue plate) destroy the which-way information. Fiber couplers are labelled with D1-D4, indicating the detectors with which they are connected in the data analysis. Multimode fibers (MMF) guide the photons from the polarization analyzers to the SPADs.
The experiment is accessed from the stationary laboratory via a Wifi-antenna. The crate is placed on a rotating table that is driven by a three-phase motor.
b) Rotational induced unitary transformation on an entangled Bell-state with a Sagnac interferometer. In our experiment the BS is replaced by a PBS and the mirrors are replaced by a fiber coil. The HV-photon pair is transformed by a half-wave plate turned at $22.5^\circ$ to a superposition of two H photons and two V photons. The horizontal (vertical) polarized part of the superposition is transmitted (reflected) at the PBS and thus co-moving (counter-moving) with rotation. In this configuration the phase of the Bell-state corresponds to the Sagnac phase. As a result, we can adjust the phase of our Bell-state just via the angular velocity $\Omega$.
For a single-photon experiment in our Sagnac interferometer, we just place a linear polarizer (POL) behind the first fiber coupler.
c) Illustration of the Sagnac effect. Mirrors form a closed and spatially overlapping path of the output ports B and C of the beamsplitter (BS). The whole setup rotates around an axis perpendicular to the enclosed area $A$ with the angular velocity $\Omega$. The incoming beam at A is divided into a counter- and co-propagating beam at the beamsplitter. The co- and counter-propagating beams are superimposed at the output port D. The phase difference, the so-called Sagnac phase $\varphi$ is linearly dependent on the angular velocity $\Omega$.
}
\label{fig:setup}
\end{figure*}

\noindent\textbf{Compact experimental setup of our Sagnac-interferometer for Bell-state phase manipulation.}\ 
Our compact setup consists of a rugged crate (54 cm long, 37 cm wide and 48 cm high) with 4 levels as depicted in Fig.~\ref{fig:setup} a). This crate is placed on top of a rotating table, driven by a three-phase motor. 
For photon pair generation, we use a spontaneous parametric down conversion (SPDC) based source in the first level of the 4-level crate. We pump a periodically poled type-II KTP crystal with a continuous wave laser diode with a central wavelength of 405~nm.
In the crystal, a pump laser photon is converted into two orthogonally polarized photons $\ket{\mathrm{H,V}}$ via SPDC, each with a wavelength of $\lambda = 810$~nm. The crystal is heated at a constant temperature of $T = 37.85 \pm 0.01^\circ \mathrm{C}$, at which the photon pair is degenerate. Two dichroic mirrors and an interference filter (IF) seperate the SPDC photons from the pump photons. For the walk-off compensation we deploy two birefringent YVO$_{4}$ crystals and a 1.3~m long polarization maintaining single mode fiber that guides the $\ket{\mathrm{H,V}}$ photon pairs to the Sagnac interferometer in the third level. 
A power supply in the second level provides electrical power for the electrical devices within the crate for $\sim 40$~minutes. In the third level, there is the polarizing beamsplitter (PBS) of the Sagnac interferometer and the polarization analyzers of Alice and Bob. The fiber coil of the Sagnac interferometer is built on top of the polarization analyzers. After the fiber coil, 50 \% of the photon pairs are spatially separated by a 50:50 beamsplitter (BS). The polarization analyzers of Alice and Bob are in the outputs of this beamsplitter. After passing the polarizing beamsplitters in the analyzers of Alice and Bob, the photons are coupled in multimode fibers which guide them to avalanche photo diodes (APD). The APDs are placed in the corners of the third level.
The time-tagging module (TTM) in the bottom writes lists with the arrival times of the photons. A software calculates the cross-correlation between the time-tag lists of the detectors and corrects for the time offsets between the detectors and thus identifies coincidences. 
With a Wifi-antenna we can access the otherwise autonomous rotating laboratory. On top of the third level we built the fiber coil inside a box made out of polysterene. \\[-2mm]

\noindent The \textbf{Sagnac effect}\
describes the phase shift of two light waves induced by rotational motion as depicted in Fig.~\ref{fig:setup} c). Mirrors form an optical closed loop of the outputs B and C of the BS. 
50 \% of the impinging beam is reflected and the other 50 \% is transmitted.
When the setup is rotating around an axis perpendicular to the enclosed area $A$, the reflected beam is co-moving while the transmitted beam is counter-moving. After passing through the interferometer both beams are superimposed at the output D. Since the co-moving beam has a longer optical path until it reaches the BS while the counter-moving beam experiences a shorter optical path, a phase shift (Sagnac phase) is induced.
This phase shift $\varphi_{\mathrm{s}}$ is linearly dependent on the angular velocity $\Omega$, as can be seen from the following equation:
\begin{equation}
\varphi_{\mathrm{s}} = \frac{8 \pi A}{\lambda c} \cdot \Omega = \frac{4 \pi L r}{\lambda c} \cdot \Omega = S_{\mathrm{T}} \cdot \Omega
\end{equation}
where $A$ is the enclosed area, $\lambda$ is the central wavelength of the photons and $c$ is the speed of light. In our experiment, a fiber coil connects B and C of the BS. Then $L$ is the fibre length, $r$ is the radius of the fibre coil. $S_{\mathrm{T}}$ denotes the Sagnac scale factor. 
\\[-2mm]

\noindent Interestingly, the Sagnac phase is not affected by the position of the rotational axis  (it is also possible that the rotation axis is outside the enclosed area) nor is it affected by the shape of the enclosed area $A$ nor is it influenced by the refractive index of the glass fiber \cite{Post_67}. \\[-2mm]

\noindent \textbf{Sagnac interferometer for phase manipulation of a Bell state}.\
In the following we describe based on Fig.~\ref{fig:setup} b), how we generate a $\ket{\phi}$-state and how we use rotational motion in order to adjust/manipulate the phase of the $\ket{\phi}$-state: \\
The source in Fig.~\ref{fig:setup} a) in the first level of the crate provides orthogonally polarized photon pairs in the same spatial mode, which are described by the following Fock-state:
\begin{equation}
\ket{\Psi^{\mathrm{Source}}} = \ket{\mathrm{H,V}} = \ket{1_{\mathrm{H}},1_{\mathrm{V}}}
\label{eq:}
\end{equation}
where one photon is horizontally (H) and the other one is vertically (V) polarized. 
A half-wave plate (HWP) set at $22.5^\circ$ in front of the Sagnac interferometer turns the $\ket{\mathrm{H,V}}$-photon pair in the DA-basis 
\begin{equation}
\hat{HWP}_{@ 22.5^\circ}\ket{\Psi^{\mathrm{Source}}}  \rightarrow \ket{\mathrm{D,A}},
\label{eq:}
\end{equation}
which corresponds in turn to the following superposition:
\begin{equation}
\ket{\Psi} = \frac{1}{\sqrt{2}} \left(  \ket{\mathrm{H,H}} - \ket{\mathrm{V,V}} \right). 
\label{eq:}
\end{equation}
At the polarizing beamsplitter (PBS) the $\ket{\mathrm{H,H}}$ part of the superposition is transmitted and is thus co-moving with the rotation, while the $\ket{\mathrm{V,V}}$ part of the superposition is reflected and thus counter-moving to the rotation. \\
The dependence of the Sagnac phase on the photon number of this state is described by applying the phase-shifting operator $\hat{U}(\varphi) = \mathrm{e}^{i\hat{N}\varphi}$ on our state, where $\hat{N}$ is the photon number operator for which $\hat{N} = \hat{a}^{\dag} \hat{a}$ and $N=2$. After passing through the fiber coil, the photons are superimposed at the PBS output D and the state reads:
\begin{equation}
\ket{\Psi^{\mathrm{SI}}} = \hat{U} \ket{\Psi} = \frac{1}{\sqrt{2}}( \ket{\mathrm{H,H}}_{\mathrm{T}}  - \mathrm{e}^{iN(\varphi_{\mathrm{s}}+o)} \ket{\mathrm{V,V}}_{\mathrm{R}} )
\label{eq:}
\end{equation}
where the subscript T stands for 'transmitted' at the PBS and R stands for 'reflected' at the PBS.
$o$ is an offset resulting from the birefringence in the fiber coil, as H- and V-polarized photons experience a slightly different refractive index within the fiber coil.
We investigate the photon pairs that are separated by the beamsplitter (BS), where one photon is measured by Alice and one photon is measured by Bob. In post-selection we get the state:
\begin{equation}
\ket{\Psi^{\mathrm{SI}}} =  \frac{1}{\sqrt{2}}( \ket{\mathrm{H_{\mathrm{A}},H_{\mathrm{B}}}}  - \mathrm{e}^{i2(\varphi_{\mathrm{s}}+o)} \ket{\mathrm{V_{\mathrm{A}},V_{\mathrm{B}}}} ).
\label{eq:}
\end{equation} 
Thus, the phase of the state depends on the Sagnac phase $\varphi_{s}$. \\[-2mm]

\noindent \textbf{Detection of the Bell-state and entanglement.} The detection of polarized photons takes place in the third level as depicted in Fig.~\ref{fig:setup} a). 
For observing interference in the coincidence measurements, the half-wave plates in front of the polarization analyzers of Alice and Bob are set at 22.5$^\circ$ in order to destroy the which-way information of whether the photon pair took the co- or counter-rotating path through the Sagnac interferometer.
For the coincidences between detectors of Alice and Bob which measure orthogonal polarization we calculate the probability:
\begin{equation}
\begin{split}
P_{\phi^-} &= \vert  \braket{\phi^- \vert \Psi^{\mathrm{SI}}}   \vert ^2 \\     
&= 
\vert
\frac{1}{\sqrt{2}} 
\left(  \bra{\mathrm{D_{\mathrm{A}},A_{\mathrm{B}}}}  
+ \bra{\mathrm{A_{\mathrm{A}},D_{\mathrm{B}}}}
\right) \frac{1}{\sqrt{2}}( \ket{\mathrm{H_{\mathrm{A}},H_{\mathrm{B}}}} \\
&- \mathrm{e}^{i2(\varphi_{\mathrm{s}}+o)} \ket{\mathrm{V_{\mathrm{A}},V_{\mathrm{B}}}} )
\vert ^2 \\
&= \mathrm{cos}^2(\varphi_{\mathrm{s}} + o)
\label{eq:Prob_Phi_min}
\end{split}
\end{equation}

For the coincidences between detectors of Alice and Bob which measure the same polarization we calculate the probability:
\begin{equation}
\begin{split}
P_{\phi^+} &= \vert  \braket{\phi^+ \vert \Psi^{\mathrm{SI}}}   \vert ^2 \\     
&= 
\vert
\frac{1}{\sqrt{2}} 
\left(  \bra{\mathrm{D_{\mathrm{A}},D_{\mathrm{B}}}}  
+ \bra{\mathrm{A_{\mathrm{A}},A_{\mathrm{B}}}}
\right) \frac{1}{\sqrt{2}}( \ket{\mathrm{H_{\mathrm{A}},H_{\mathrm{B}}}} \\
&- \mathrm{e}^{i2(\varphi_{\mathrm{s}}+o)} \ket{\mathrm{V_{\mathrm{A}},V_{\mathrm{B}}}} )
\vert ^2 \\
&= \mathrm{sin}^2(\varphi_{\mathrm{s}} + o)
\label{eq:Prob_Phi_plus}
\end{split}
\end{equation}
 \\[-2mm]

\begin{figure*}[t]
\centering
\includegraphics[width=0.99\textwidth]{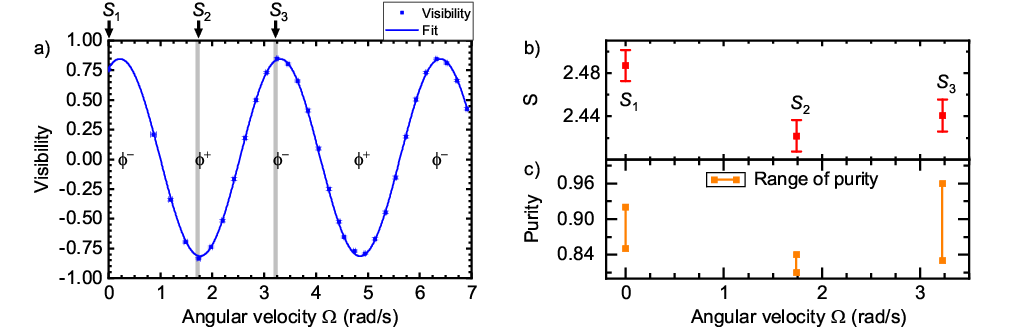}
\caption{\textbf{Visibility, CHSH-Bell parameter $S$ and the range of the purity.} 
a) Visibility of the coincidence measurements versus the angular velocity between the polarization analyzers of Alice and Bob. The integration time of the coincidences was 40~s long. A least-square fit provides a maximum visibility of $V = 83.05 \pm 0.29 \%$. Vertical error bars are calculated with gaussian error propagation. For the coincidences we assume a Poissonian distribution.
$S_1$, $S_2$ and $S_3$ label the points where we measured the CHSH quantity $S$. In the first crest position at $\sim 0.21$~rad/s the photon pairs are in a $\ket{\phi^{-}}$-state. At the trough position at $\sim 1.74$~rad/s, the photon pairs are described by a $\ket{\phi^{+}}$-state.
b) shows the measured $S$-values. ($S$-value measurements at 0 rad/s, 1.74 rad/s, and 3.23 rad/s). The lower part of the plot in c) shows the range of the purity which we gain from the partial quantum state tomography.
}
\label{fig:Omega_S_Purity}
\end{figure*}

\noindent\textbf{Results}.
Fig.~\ref{fig:Omega_S_Purity} shows the results of the rotation measurements. In a) we can see the visibility dependent on the angular velocity $\Omega$, while b) and c) show the Bell-CHSH parameter $S$ and the range of the purity, respectively at the points $S_1$, $S_2$ and $S_3$ in Fig.~\ref{fig:Omega_S_Purity} a).
By using Eq.~\ref{eq:Prob_Phi_min}~and~\ref{eq:Prob_Phi_plus} (see Fig.~\ref{fig:results_coincidences_all} in the Appendix for the coincidence measurements between the analyzers of Alice and Bob) we calculate the expected visibility, 
which is the fit function $V(\Omega) = A \cdot \mathrm{cos}(2 S_{\mathrm{T}} \Omega + o) + D$ in the plot (for the derivation see Eq.~\ref{eq:visibility} in the Appendix).
The measured data fit very well to the expected visibility in Eq.~\ref{eq:visibility} and show a lateral offset $o$. We attribute this lateral offset of the data to birefringence in the fiber coil, since H- and V- polarized photons see a slightly different refractive index (This lateral offset is also observable in the coincidence measurements in Fig.~\ref{fig:results_coincidences_all}). 
In the first crest position of Fig.~\ref{fig:Omega_S_Purity} a) at $\sim 0.21$~rad/s the photons are entangled in a $\ket{\phi^{-}}$-state. Then, in the next trough position at $\sim 1.74$~rad/s labelled with $S_2$ the photons are entangled in a $\ket{\phi^{+}}$-state, and so on. \\
\\
\noindent In a next step, we measure the Bell-CHSH parameter $S$ in order to provide a further figure of merit that proves that our photon pairs are entangled in the first three maximum and minimum positions of our coincidence measurements in Fig.~\ref{fig:results_coincidences_all} the corresponding states as indicated in Fig.~\ref{fig:Omega_S_Purity} a). 
With the coincidences between Alice and Bob $C(\alpha,\beta)$ we calculate the correlation function: 
\begin{equation}
E(\alpha, \beta) = \frac{C(\alpha, \beta)+C(\overline{\alpha}, \overline{\beta})-C(\alpha, \overline{\beta})-C(\overline{\alpha}, \beta)}{C(\alpha, \beta)+C(\overline{\alpha}, \overline{\beta})+C(\alpha, \overline{\beta})+C(\overline{\alpha}, \beta)},
\label{eq:corr_fctn}
\end{equation}
where $\alpha \perp \overline{\alpha}$ and $\beta \perp \overline{\beta}$.
Using Eq. \ref{eq:corr_fctn}
and the three other correlation functions $E(\alpha, \beta')$, $E(\alpha', \beta)$ and $E(\alpha', \beta')$, we finally calculate the Bell-CHSH parameter
\begin{equation}
S = \left| E(\alpha, \beta) - E(\alpha, \beta') \right| + \left| E(\alpha', \beta) + E(\alpha', \beta') \right|.
\end{equation}
For a maximum violation of the $\ket{\phi^-}$-Bell state ($\ket{\phi^+}$-Bell state) we use the following angles:
Alice $\alpha = 0^\circ$ and $\alpha' = -45^\circ$ 
(Alice $\alpha = 0^\circ$ and $\alpha' = -45^\circ$)
Bob: $\beta = 22.5^\circ$ and $\beta' = 67.5^\circ$ 
(Bob: $\beta = -22.5^\circ$ and $\beta' = -67.5^\circ$).
\\[-2mm]

In the first maximum position of the visibility $V(\Omega)$ in Fig.~\ref{fig:Omega_S_Purity}a) our photon pairs can be described by a $\ket{\phi^-}$-Bell state. Due to a slight offset of $\sim 0.21$~rad/s, the maximum position is not at zero. 
A subsequent measurement of $S$ after the coincidence measurements while the interferometer is resting, and thus $\sim 0.21$~rad/s away from the maximum and minimum position, yields $S_1 = 2.4869 \pm 0.0142$.
Here, we violate the classical bound by 34.3 standard deviations. For the error of the coincidences we assume a Poissonian distribution and calculate the standard deviation via Gaussian error propagation. Although, we are not exactly at the maximum position, but have the aforementioned offset, our photon pairs are still entangled. \\
On the next day, after charging the power supply, we measured $S$ at $\Omega = 1.74$~rad/s and $\Omega = 3.23$~rad/s. 
At these angular velocities, we measured $S_2 = 2.4219 \pm 0.0147$ and $S_3 = 2.4407 \pm 0.0149$, respectively.
The integration time for the coincidences is 80 s long. In Fig.~\ref{fig:Omega_S_Purity} and \ref{fig:results_coincidences_all}, they are labelled as $S_{2}$ and $S_{3}$, respectively. As can be seen in Fig.~\ref{fig:results_coincidences_all}, $S_{1}$ and $S_{3}$ are the Bell-CHSH parameters in the first and third maximum position, in which our photon pairs can be described by a $\ket{\phi^-}$-Bell state. In the second maximum position, for which we measure $S_{2}$, we have a $\ket{\phi^+}$-Bell state. The results of these measurements are listed in 
Fig.~\ref{fig:Omega_S_Purity} b). All of them violate the classical bound of $S_{\mathrm{classical}} = 2$ of at least 28.6 standard deviations, thus proving that our states are entangled in the corresponding state.
\\[-2mm]

Finally, we perform a partial quantum state tomography with the data of the coincidence measurements for the Bell-CHSH parameters $S_1$, $S_2$ and $S_3$. With this data we determine the diagonal elements of the density matrices and an upper bound for the off-diagonal elements. We calculate a range for the purity of our Bell states with the upper bound.  Fig.~\ref{fig:Omega_S_Purity} c) shows the range of the purity for the quantum state measured at $\Omega = 0, \, 1.74$ and $3.23$~rad/s. The density matrices can be found in Fig.~\ref{fig:density_matrix_F_1}, \ref{fig:density_matrix_F_2} and \ref{fig:density_matrix_F_3} in the Appendix. \\
From the diagonal elements $\rho_{00}$ and $\rho_{11}$ of the density matrices we see that the states are not perfectly symmetric, with more photon pairs exhibiting vertical polarization. 
\begin{equation}
\ket{\Psi^{\mathrm{SI}}} =  \sqrt{\rho_{00}}\ket{\mathrm{H_{\mathrm{A}},H_{\mathrm{B}}}}  - \sqrt{\rho_{11}} \mathrm{e}^{i2(\varphi_{\mathrm{s}}+o)} \ket{\mathrm{V_{\mathrm{A}},V_{\mathrm{B}}}} 
\label{eq:}
\end{equation} 
Since $\rho_{00} < \rho_{11}$ for $S_1$, $S_2$ and $S_3$, the $\ket{\phi^{\pm}}$-states are asymmetric, which explains the difference of up to 0.4 between the perfect Bell violation of $S = 2.82$ and the measured $S$-values, despite the high purity. \\
\\
\noindent\textbf{Summary and conclusion}.\ 
We built a Sagnac interferometer for phase manipulation of an entangled Bell-state and we successfully demonstrated that we can switch between a $\ket{\phi^{-}}$ and a $\ket{\phi^{+}}$ state just via changing the angular velocity of this interferometer. Within the range of $\sim 7$~rad/s we could show that we can switch between these two states for four times. 
Notably, we achieved this phase manipulation via non-inertial motion. 
We proved polarization-entanglement of the $\ket{\phi^{-}}$ and $\ket{\phi^{+}}$ states with the Bell-CHSH parameter $S$, providing a mostly device independent proof of entanglement (modulo the obvious loopholes of Bell experiments). An interesting note is that while the experiment follows the expected sinusoidal curve for the correlations, the actual Bell value attained deviated from the maximum, which turned out to indicate a non-maximally entangled state more than decoherence. 
Indeed, using the coincidences from the Bell measurement we conducted a partial quantum state tomography, with which we gained the diagonal elements of the density matrix of the states. Thereby, we could determine an upper bound for the off-diagonal elements and a range for the purity. Observing the experimental Bell violation it is interesting to note that while at first, the Bell value actually decreases (which would indicate decoherence due to non-stabilities of the interferometric setup), even faster rotations recover higher Bell values, indicating that rotational stability of the setup was not the key issue in reduced Bell violation and the setup is indeed very stable. 
\\[-3mm] 


\bibliographystyle{apsrev4-1fixed_with_article_titles_full_names_new}
\bibliography{references}


\vspace*{3mm}
\noindent\textbf{Acknowledgments}.\ 
We thank Roland Blach for building the table with the rotating platform. 
We acknowledge the Austrian Academy of Sciences in cooperation with the FhG ICON-Program "Integrated Photonic Solutions for Quantum Technologies (InteQuant)".
\\

\noindent\textbf{Author contributions}.\ 
M.F. built the photon-pair source and conceived the experiment. J.B. built the setup, conducted the experiment and data analysis. J.B. and M.H. wrote the manuscript. R.U. and M.H. supervised the experiment. All authors contributed to discussions of the results and the manuscript.


\newpage
\clearpage
\begin{widetext}

\hypertarget{sec:appendix}
\appendix

\section*{Appendix: Supplemental Information}

\renewcommand{\thesubsubsection}{A.\Roman{subsection}.\arabic{subsubsection}}
\renewcommand{\thesubsection}{A.\Roman{subsection}}
\renewcommand{\thesection}{}
\setcounter{equation}{0}
\numberwithin{equation}{section}
\setcounter{figure}{0}
\renewcommand{\theequation}{A.\arabic{equation}}
\renewcommand{\thefigure}{A.\arabic{figure}}

\subsection{Additional information to the measurement}
\noindent To the measurement of the angular velocity during coincidence measurements (in Fig.~\ref{fig:Omega_S_Purity}a) and Fig.~\ref{fig:results_coincidences_all}): \\
For the measurement of the angular velocity $\Omega$ we use the VibraScout6D sensor by Dytran Instruments. During the course of the measurements the data show sudden rectangular shaped jumps in between which don't map the actual angular velocity. For the calculation of the average of the angular velocity, we cut out these data. \\
\\
\noindent To the measurement of the angular velocity during the coincidence measurements of the CHSH-Bell quantity $S_2$ (in Fig.~\ref{fig:Omega_S_Purity}): \\
During the measurement of the coincidences for the expectation value of
$E(\alpha,\beta')$ for the calculation of $S_2$ (at $\Omega = 1.74$~rad/s) the 
Gyrosensor was not working properly. The data of the angular velocity
show a course that is not in accordance with the constant angular velocity depicted
on the segment display of the frequency drive of the motor. Nevertheless, we used 
the acquired data for the calculation of $S_2$. 

\subsection{Single photon interference}
\noindent In a subsequent measurement, we placed a polarization filter directly behind the PMSMF (blue optical element in Fig.~\ref{fig:setup} b)), thus reducing the two-photon state to a single-photon state. As for the two-photon measurements, the half-wave plate after the PMSMF is turned at an angle of $22.5^\circ$. A horizontal (vertical) polarized single photon described by the ket vector $\ket{\mathrm{H}}$ ($\ket{\mathrm{V}}$) is thus turned to a diagonal (antidiagonal) polarized photon $\ket{\mathrm{D}}$ ($\ket{\mathrm{A}}$) which is in turn a superposition of a horizontal and vertical polarization:
\begin{equation}
    \ket{\mathrm{D}} = \frac{1}{\sqrt{2}} ( \ket{\mathrm{H}} + e^{i\hat{N}(\varphi_{\mathrm{s}}+o)}\ket{\mathrm{V}} )
    \label{eq:single_photon_state}
\end{equation}
where $\hat{N}$ is the photon number operator, for which we insert $N = 1$ in this part of the experiment. As for the two-photon state (described in the main text), the horizontal polarized part of the superposition is transmitted, while the vertical part of the transmission is reflected at the PBS in Fig.~\ref{fig:setup} b). When the setup is rotating, the state described in Eq.~\ref{eq:single_photon_state} acquires the Sagnac phase $\varphi_{\mathrm{s}}$. 

\begin{figure*}[h] 
\includegraphics[width=0.48\textwidth]{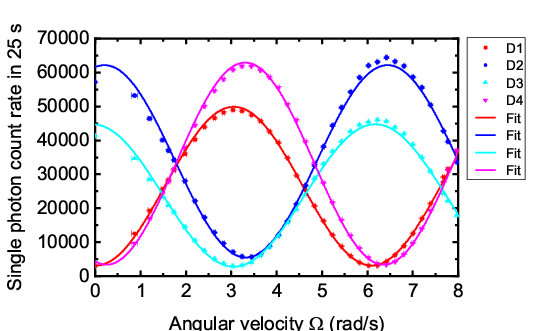}
\caption{\textbf{Results of the single photon measurements} with only one photon in the Sagnac interferometer. Single photon count rates integrated over 25 s vs. angular velocity $\Omega$ for all detectors of Alice and Bob. The single photon count rates of detector D1 and D4 are proportional to $\mathrm{sin}^2(S_{\mathrm{T}} \cdot \Omega)$, whereas the single photon count rates of D2 and D4 show an interference pattern proportional to $\mathrm{cos}^2(S_{\mathrm{T}} \cdot \Omega)$, both with an offset along the x-axis, which we attribute to birefringence in the fiber coil. 
For the error bars of the single photon count rates we assume a Poissonian distribution.
Error bars for the angular velocity are the minimum and maximum angular velocity.
}
\label{fig:results_singles_all}
\end{figure*}

\noindent We measured the single photon count rates for each detector of Alice and Bob, shown in Fig.~\ref{fig:results_singles_all}. The figure shows a course of the single photon count rates that is proportional to $\mathrm{sin}^2(S_{\mathrm{T}}\cdot \Omega)$ and $\mathrm{cos}^2(S_{\mathrm{T}}\cdot \Omega)$. Thus the single photon count rates show a similar course as the coincidence measurements in Fig.~\ref{fig:results_coincidences_all}, however with half the periodicity. The valleys in the fringes show a doubled distance compared to the coincidence measurements in Fig.~\ref{fig:results_coincidences_all}. 
This is explained by the photon number operator $\hat{N}$ in Eq.~\ref{eq:single_photon_state}.
A least-square fit provides the scaling factor $S_{\mathrm{T}}$, the offset $o$ and the visibility $V$ for each single photon count rate dependent on the angular velocity. They are listed in table \ref{table:Fit_data_Singles}. 
 
\begin{table}[H]
\centering
\begin{tabular}{ |c|c|c|c| } 
 \hline
 Det.  & $S_{\mathrm{T}}$ & $o$ & $V$ \\
 \hline
 1 & $0.5169 \pm 0.0012$ & $-0.0108 \pm 0.0064$ & $88.23 \pm 0.35 \%$ \\ 
 2 & $0.5035 \pm 0.0023$ & $-0.1048 \pm 0.0103$ & $83.79 \pm 0.64 \%$ \\ 
 3 & $0.5041 \pm 0.0019$ & $0.0272 \pm 0.0089$ & $88.33 \pm 0.58 \%$ \\ 
 4 & $0.5142 \pm 0.0013$ & $-0.1310 \pm 0.0069$ & $89.87 \pm 0.41 \%$ \\ 
 \hline
\end{tabular}
\caption{\textbf{Fit data}  of the Sagnac scaling factor $S_{\mathrm{T}}$, the offsets $o$ and the visibilities $V$ for the single photon measurement from Fig.~\ref{fig:results_singles_all}.}. 
\label{table:Fit_data_Singles}
\end{table}

\subsection{Coincidence measurements of two photon interference}

\noindent Fig.~\ref{fig:results_coincidences_all} shows the results of the coincidence measurements dependent on the angular velocity $\Omega$ between detectors of Alice and Bob.
The magenta and cyan data points are the coincidences between detectors which measure the same polarization, while the red and blue data points show the coincidences of detector combinations between Alice and Bob with orthogonal polarization.
We measure coincidences in the range between $\Omega = 0 \, \mathrm{to} \, 7 $~rad/s.
Least square fits of the form 

\begin{equation}
F_{\mathrm{s}}(\Omega) = A \cdot \mathrm{sin}^2(S_{\mathrm{T}} \cdot \Omega + o) + D
\end{equation}

\begin{equation}
F_{\mathrm{c}}(\Omega) = A \cdot \mathrm{cos}^2(S_{\mathrm{T}} \cdot \Omega + o) + D
\end{equation}
provide the following fit data, listed in table
\ref{table:Fit_data}. Detector pairs between Alice and Bob with same (orthogonal) polarization are proportional to $\mathrm{sin}^2(S_{\mathrm{T}} \cdot \Omega + o)$ ($\mathrm{cos}^2(S_{\mathrm{T}} \cdot \Omega + o)$) as calculated in Eq.~\ref{eq:Prob_Phi_min}~and~\ref{eq:Prob_Phi_plus}. From the fits we extract visibilities for each channel combination in Fig.~\ref{fig:results_coincidences_all}, listed in column 4 of the table~\ref{table:Fit_data}. 

\begin{figure*}[ht!] 
\includegraphics[width=0.48\textwidth]{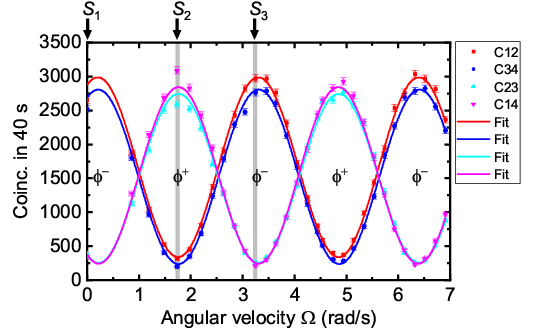}
\caption{\textbf{Results of the coincidence measurements} Coincidences integrated over 40 s vs. angular velocity $\Omega$ for all detector combinations between Alice and Bob. C12 in the legend stands for coincidences between detector 1 and 2. Analyzer combinations between Alice and Bob with the same polarization show an interference pattern proportional to $\mathrm{sin}^2(S \cdot \Omega)$, whereas polarizer combinations with orthogonal polarizations show an interference pattern proportional to $\mathrm{cos}^2(S \cdot \Omega)$, both with an offset along the x-axis, which we attribute to birefringence in the fiber. $S_1$, $S_2$ and $S_3$ label the angular velocities at which we conducted a CHSH Bell test measurement. In the first position of the crests an troughs at $\sim 0.21$~rad/s the photon pairs are in a $\ket{\phi^{-}}$-state. At the next crest and trough position at $\sim 1.74$~rad/s labelled with $S_2$ photons are entangled in a $\ket{\phi^{+}}$-state and so on.   
For the error bars of the coincidences we assume Poissonian distribution.
Error bars for the angular velocity correspond to the minimum and maximum angular velocity.
}
\label{fig:results_coincidences_all}
\end{figure*}

\begin{table}[h!]
\centering
\begin{tabular}{ |c|c|c|c| } 
 \hline
 Det. comb. & $S_{\mathrm{T}}$ & $o$ & $V$ \\
 \hline
 1 - 2 & $1.015 \pm 0.003$ & $-0.2111 \pm 0.0126$ & $79.69 \pm 0.71 \%$ \\ 
 3 - 4 & $1.014 \pm 0.003$ & $-0.2181 \pm 0.0119$ & $84.44 \pm 0.66 \%$ \\ 
 2 - 3 & $1.017 \pm 0.003$ & $-0.2247 \pm 0.0131$ & $83.12 \pm 0.73 \%$ \\ 
 1 - 4 & $1.02 \pm 0.003$ & $-0.229 \pm 0.0140$ & $84.27 \pm 0.78 \%$ \\ 
 \hline
\end{tabular}
\caption{\textbf{Fit data}  of the Sagnac scaling factor $S_{\mathrm{T}}$, the offsets $o$ and the visibilities $V$ for the two photon measurement in Fig.~\ref{fig:results_coincidences_all}.}
\label{table:Fit_data}
\end{table}  

\noindent Our fiber coil has a radius of $r = 7.8$~cm. With the provided Sagnac scaling factors we calculate a fiber length of $L = 251.46 \pm 3.31$~m ($L = 251.21 \pm 3.31$, $L = 251.95 \pm 3.31$, $L = 251.70 \pm 3.31$). \\
In the first maximum of the coincidences of detector pair 1 and 2 (3 and 4) in Fig.~\ref{fig:results_coincidences_all} our photon pairs are in a $\ket{\phi^{-}}$-state. At a higher angular velocity of $\sim 1.74$~rad/s the coincidences reach a minimum in which the photon pairs are entangled in a $\ket{\phi^{+}}$-state. In the next maximum and minimum position at an angular velocity of $\sim 3.23$~rad/s the photon pairs are again in a $\ket{\phi^{-}}$-state.
This coincidence measurements show the successful implementation of the phase manipulation of a polarization-entangled photon pair by means of a Sagnac interferometer, in which we change the state just by increasing the angular velocity and thus the state of non-inertial motion. \\
\\
By using the coincidence measurements from Fig.~\ref{fig:results_coincidences_all} we calculate the visibility shown in Fig.~\ref{fig:Omega_S_Purity}.
We calculate the formula for the fitfunction of the visibility of the coincidence measurements between the polarization analyzers of Alice and Bob from Fig.~\ref{fig:Omega_S_Purity} in the following way:
\begin{equation}
V(\varphi_{\mathrm{s}}) = \frac{C_{\mathrm{DD}} + C_{\mathrm{AA}} - C_{\mathrm{DA}} - C_{\mathrm{AD}}}{C_{\mathrm{DD}} + C_{\mathrm{AA}} + C_{\mathrm{DA}} + C_{\mathrm{AD}}} =
\frac{2 \cdot \mathrm{cos}^2(\varphi_{\mathrm{s}}) - 2 \cdot \mathrm{sin}^2(\varphi_{\mathrm{s}})}{2 \cdot \mathrm{cos}^2(\varphi_{\mathrm{s}}) + 2 \cdot \mathrm{sin}^2(\varphi_{\mathrm{s}})} =
\mathrm{cos}(2\varphi_{\mathrm{s}}),
\label{eq:visibility}
\end{equation}
where $C_{\mathrm{DD}}$ are the coincidences between the detectors of Alice and Bob, which measure both diagonal (D) polarized photons. The subscript A stands for an anti-diagonal polarization. The first subscript denotes the polarization of Alice's detector and the second subscript denotes the polarization of Bob's detector. A fit function of the form
\begin{equation}
F_V(\Omega) = A \cdot \mathrm{cos}(2\cdot S_{\mathrm{T}} \cdot \Omega + o) +D
\label{eq:visibility_fit}
\end{equation}
is applied to the measured data. The least-square fit provides: $A=0.8305\pm0.0029$, $B = 1.017\pm0.001$, $C =-0.4476\pm0.0111$, $D =0.0152\pm0.0023$.

\subsection{Partial quantum state tomography}
\noindent This section provides the bar graphs with the density matrices of the Bell-states at the angular velocities of $\Omega = 0, 1.74$ and $3.23$~rad/s (Fig.~\ref{fig:density_matrix_F_1}, \ref{fig:density_matrix_F_2} and \ref{fig:density_matrix_F_3}). The Bell-states in this experiment are mathematically described by 4x4 density matrices, of the shape:
\begin{equation}
\hat{\rho}_{\mathrm{exp}} = \left( \begin{array}{rrrr} 
\rho_{\mathrm{00}} & \rho_{\mathrm{0001}} & \rho_{\mathrm{0010}} & \rho_{\mathrm{0011}} \\ 
\rho_{\mathrm{0100}} & \rho_{\mathrm{01}} & \rho_{\mathrm{0110}} & \rho_{\mathrm{0111}} \\ 
\rho_{\mathrm{1000}} & \rho_{\mathrm{1001}} & \rho_{\mathrm{10}} & \rho_{\mathrm{1011}} \\ 
\rho_{\mathrm{1100}} & \rho_{\mathrm{1101}} & \rho_{\mathrm{1110}} & \rho_{\mathrm{11}} \\ 
\end{array} \right).
\end{equation}
The diagonal elements are calculated from the coincidence measurements of the CHSH-Bell parameter $S$. By means of the Cauchy-Schwarz inequality we determine an upper bound for the off-diagonal elements:
\begin{equation}
    \vert \bra{i} \hat{\rho} \ket{j} \vert \leq \sqrt{  \bra{i} \hat{\rho} \ket{i} \bra{j} \hat{\rho} \ket{j} },
    \label{eq:CauchySchwarz}
\end{equation}
where $i$ refers to the row index and $j$ refers to the column index. For the calculation of the upper and lower bounds of the matrix elements $\rho_{\mathrm{0011}}$ and $\rho_{\mathrm{1100}}$ we use the CHSH-Bell inequality and equate it with the experimental value $S_{\mathrm{exp}}$ from Fig.~\ref{fig:Omega_S_Purity}. For the $\ket{\phi^-}$-state we use the CHSH-Bell inequality
\begin{equation}
    S_{\mathrm{exp}} \stackrel{!}{=} \frac{2}{\sqrt2} \left(    \mathrm{Tr}[\sigma_{z} \otimes \sigma_{z} \cdot \rho] - \mathrm{Tr}[\sigma_{x} \otimes \sigma_{x} \cdot \rho]   \right),
\end{equation}
and for the $\ket{\phi^+}$-state we use the CHSH-Bell inequality
\begin{equation}
    S_{\mathrm{exp}} \stackrel{!}{=} \frac{2}{\sqrt2} \left(    \mathrm{Tr}[\sigma_{z} \otimes \sigma_{z} \cdot \rho] + \mathrm{Tr}[\sigma_{x} \otimes \sigma_{x} \cdot \rho]   \right).
\end{equation}

\noindent The next three graphs in Fig.~\ref{fig:density_matrix_F_1}, \ref{fig:density_matrix_F_2} and \ref{fig:density_matrix_F_3}, show the results of this calculation. The left bar graphs show the upper bound matrices $\rho_{\mathrm{exp,ub}}$, while the right bar graphs show the lower bound matrices $\rho_{\mathrm{exp,lb}}$.

\begin{figure*}[ht!]
\centering
\includegraphics[width=0.48\textwidth]{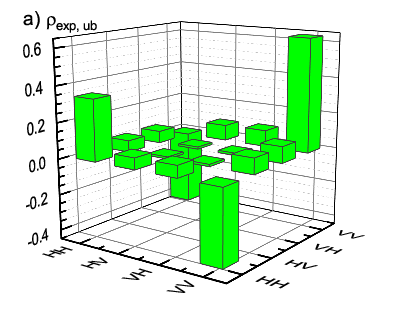}
\includegraphics[width=0.48\textwidth]{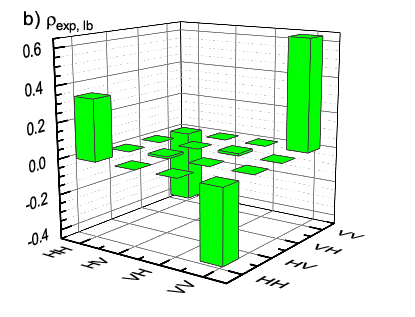}
\vspace*{-6mm}
\caption{\textbf{Density matrices} 
of the $\ket{\phi^{-}}$-state when the interferometer is not rotating. The left side in a) shows the upper bound matrix, while the right side in b) shows the lower bound matrix. With both matrices we calculate the range of the purity of the state (see Fig. \ref{fig:Omega_S_Purity}). The matrix elements $\rho_{00}$ and $\rho_{11}$ differ by almost a factor of 2 and thus the state is asymmetrical.
}
\label{fig:density_matrix_F_1}
\end{figure*}

\begin{figure*}[ht!] 
\centering
\includegraphics[width=0.48\textwidth]{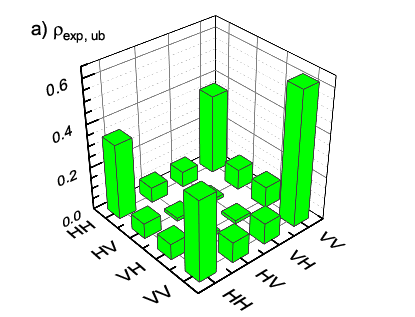}
\includegraphics[width=0.48\textwidth]{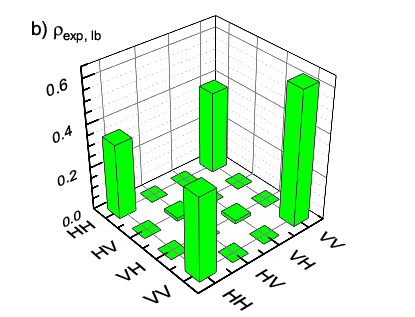}
\vspace*{-6mm}
\caption{\textbf{Density matrices} 
of the $\ket{\phi^{+}}$-state at an angular velocity of $\Omega = 1.74$~rad/s. The left side in a) shows the upper bound matrix, while the right side in b) shows the lower bound matrix. With both matrices we calculate the range of the purity of the state (see Fig. \ref{fig:Omega_S_Purity}). As in Fig. \ref{fig:density_matrix_F_1}, the matrix elements $\rho_{00}$ and $\rho_{11}$ differ by almost a factor of 2 and thus the state is asymmetrical. 
}
\label{fig:density_matrix_F_2}
\end{figure*}

\begin{figure*}[ht!] 
\centering
\includegraphics[width=0.48\textwidth]{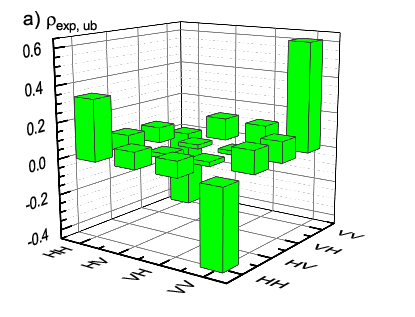}
\includegraphics[width=0.48\textwidth]{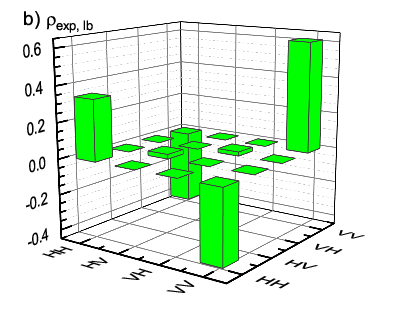}
\vspace*{-6mm}
\caption{\textbf{Density matrices} 
of the $\ket{\phi^{-}}$-state at an angular velocity of $\Omega = 3.23$~rad/s. The left side in a) shows the upper bound matrix, while the right side in b) shows the lower bound matrix. With both matrices we calculate the range of the purity of the state (see Fig. \ref{fig:Omega_S_Purity}). As in Fig. \ref{fig:density_matrix_F_1} and \ref{fig:density_matrix_F_2}, the matrix elements $\rho_{00}$ and $\rho_{11}$ differ by almost a factor of 2 and thus the state is asymmetrical. 
}
\label{fig:density_matrix_F_3}
\end{figure*}

\end{widetext}
\end{document}